\documentstyle[preprint,aps,epsf]{revtex}

\begin{document}

\title{Strange nuclear matter within Brueckner-Hartree-Fock
theory}

\author{I.\ Vida\~na, A.\ Polls, and A.\ Ramos}

\address{Departament d'Estructura i Constituents de la Mat\`eria,
Universitat de Barcelona, E-08028 Barcelona, Spain}

\author{M.\  Hjorth-Jensen}

\address{Department of Physics, University of Oslo, N-0316 Oslo,
Norway}

\author{V.G.J. Stoks}

\address{Centre for the Subatomic Structure of Matter, University
of Adelaide,
SA 5005, Australia}

\maketitle

\begin{abstract}

We have developed a formalism for microscopic
Brueckner-type calculations of dense nuclear 
matter that includes all types of
baryon--baryon interactions and allows to treat
any asymmetry on the fractions of the
different species ($n, p, \Lambda, \Sigma^-, \Sigma^0, \Sigma^+,
\Xi^-$ and $\Xi^0$).
We present results for the different single--particle potentials
focussing on situations that can
be relevant in future microscopic studies of
beta-stable neutron star matter with strangeness.
We find that both the hyperon-nucleon and hyperon-hyperon
interactions play a
non-negligible role in determining the chemical
potentials of the different species. 


\end{abstract}



\section{Introduction}
\label{sec:intro}

The properties and composition of dense matter at supranuclear
densities
determine the static and dynamical behavior of stellar matter 
\cite{glendenning92,lattimer91,cook94,pke95,prakash97}. The study
of matter at
extreme
densities and temperatures has received a renewed interest due to
the
possibility of attaining such conditions in
relativistic heavy-ion collisions at GSI, and in the near future
at CERN and Brookhaven. 

It is believed that at extremely high densities, deconfinement
will take place resulting in a transition from hadronic to quark
matter.
The transition point and its characteristics will depend
crucially on the equation of state of matter in the hadronic
phase. It is well known that the presence of strangeness, in the
form of hyperons ($\Lambda$,$\Sigma$) or mesons $(K^-)$ will
soften the equation of state and will delay the transition.
Most investigations up to date are made in the framework of the
mean field approach, either relativistic \cite{ellis95,ms96}
or non-relativistic
with effective Skyrme interactions \cite{bg97}. Microscopic
theories, on the other hand,
aim at obtaining the properties of hadrons in dense matter from
the
bare free space interaction. In this sense, Brueckner theory
was developed long time ago and successfully allowed to
understand the properties of (non-strange)
nuclear matter starting from interactions that reproduce a huge
amount
of $NN$ scattering observables. A first attempt to incorporate
strangeness
in the form of hyperons within Brueckner
theory was made in Refs. \cite{schulze1,schulze2},
latter extended to
investigations
of beta-stable nuclear matter \cite{schulze3}. A missing
ingredient in
these
works was the hyperon-hyperon ($YY$) interaction and the results
of
single--particle potentials or binding energy per baryon with a
finite amount of hyperons were simply orientative.

The recent availability
of a baryon-baryon potential \cite{sr99} covering the complete
SU(3)$\times$SU(3)
sector has allowed to incorporate the $YY$ potential in
a microscopic calculation of 
dense matter with non-zero hyperon fraction \cite{sl99}.
The incorporation of all possible baryon-baryon interactions
required the solution of the $G$-matrix equation in coupled
channels for
different strangeness sectors: $NN$ ($S=0$), $YN$ ($S=-1$),
$YY$ ($S=-2,-3$ and $-4$).
The work of Ref. \cite{sl99} concentrated mainly on isospin
saturated systems, i.e.,
systems with the same fraction of particles within the same
isospin
and strangeness multiplet: $T=1/2,S=0$ (neutrons and protons),
$T=0,S=-1$
($\Lambda$), $T=1,S=-1$ ($\Sigma^-,\Sigma^0,\Sigma^+$) and
$T=1/2,S=-2$ ($\Xi^-,\Xi^0$). In this way, the complications
associated
to different Fermi seas for each species of the same
isospin-strangeness
multiplet were avoided and the $G$-matrix in each sector was
independent on
the third component of isospin.

It is well known, however, that the presence
of electrons makes nuclear star matter to be equilibrated against the
weak $\beta$-decay reactions for neutron fractions much larger
(a factor of 10 or more) than that for protons 
\cite{bombaci91,engvik94,engvik96}.
Also, the increase of negatively
charged leptons with the baryonic density will turn into
a decrease when the appearance of
negatively charged baryons becomes energetically
more favorable. This is the case of
the $\Sigma^-$ hyperon, since neutralizing the proton charge
with $\Sigma^-$ instead of $e^-$ will remove two energetic
neutrons ($p \Sigma^- \leftrightarrow n n$) instead of one
($p e^- \leftrightarrow n$). It is clear, therefore, that a
microscopic study of $\beta$-stable nuclear matter with hyperons
requires the treatment of highly asymmetric matter, both in the
non-strange sector (protons vs. neutrons) and the hyperonic one
($\Sigma^-$ vs $\Sigma^0$ and $\Sigma^+$).
In the present paper we extend the study of Ref. \cite{sl99} to
allow for different fractions of each species. We will also
explore the effect of the recently available $YY$ interaction on 
the single--particle potential of the hyperons, a crucial
ingredient to determine the baryonic density at which the
different
hyperons appear.
Our aim is to present a thorough analysis of the
properties of the different
baryons in dense matter, taking into account their mutual
interactions. We will explore different baryonic densities and
compositions that are relevant
in the study of neutron stars.

\section{Formalism}
\label{sec:formalism}

In this section we present the formalism to obtain, in the
Brueckner--Hartree--Fock approximation, the single--particle
energies of $n$, $p$, $\Lambda$, $\Sigma^-$, $\Sigma^0$, $\Sigma^+$, $\Xi^-$ and
$\Xi^0$ embedded in an infinite system composed of different
concentrations of such 
baryons. We first construct effective baryon-baryon ($BB$)
interactions ($G$-matrices) starting from 
new realistic bare $BB$
interactions, which have become recently available
for different strangeness channels\cite{sr99}. 

\subsection{Effective $BB$ interaction}
The effective $BB$ interaction or $G$-matrix is obtained from the
bare $BB$
interaction by solving the corresponding Bethe--Goldstone
equation, which in
partial wave decomposition and using the quantum numbers of the
relative and
center--of--mass motion (RCM) reads 

\begin{eqnarray}
\lefteqn{\left\langle (B_3B_4)k''KL'' S'' (J) T,M_T\right |
      G(\omega)\left | (B_1B_2)kKLS(J)TM_T \right\rangle
      =} \hspace{2cm}\nonumber\\
  && \left\langle (B_3B_4)k''KL''S''(J) TM_T\right |
      V\left | (B_1B_2)kKLS(J)TM_T \right\rangle
      \nonumber \\
      &\quad +& {\displaystyle
      \sum_{L'}\sum_{S'}\sum_{B\widetilde{B}}\int k'^{2}dk'}
      \left\langle (B_3B_4)k''KL''S''(J)TM_T\right |
      V\left | (B\widetilde{B})k'KL'S'(J)TM_T \right\rangle
\nonumber
\\   
      && \times \frac{\overline{Q}_{B\widetilde{B}}(k',K;T,M_T)}{\omega
-\frac{K^2}{2(M_{B}+M_{\widetilde{B}})}
-
\frac{k'^2(M_{B}+M_{\widetilde{B}})}{2M_{B}M_{\widetilde{B}}}-M_{
B}-M_{\widetilde{B}}
+i{\eta}}
\nonumber \\
      && \times\left\langle (B\widetilde{B})k'KL'S'(J)TM_T\right
|
      G(\omega)\left | (B_1B_2)kKLS(J)TM_T
\right\rangle \ ,
   \label{eq:gmat}
\end{eqnarray}

The starting energy $\omega$ corresponds to
the sum of non--relativistic single--particle energies of the
interacting baryons
including their rest masses.
Note that we use kinetic energy spectrum for the intermediate
$B\widetilde{B}$
states. 
The variables $k$, $k'$, $k''$ and $L$, $L'$, $L''$ denote
relative linear momenta
and orbital momenta, respectively, while $K$ is the linear
center-of-mass momentum. 
The total angular momentum, spin, isospin and isospin projections
are denoted by $J$, $S$, $T$ and $M_T$,
respectively.
As usually, $\overline{Q}_{B\widetilde{B}}(k',K;T,M_T)$ is the angle--average of the Pauli operator
which prevents the
intermediate baryons $B$ and
$\widetilde{B}$ to be
scattered to states below their respective Fermi momenta
$k_{F}^{(B)}$ and
$k_{F}^{(\widetilde{B})}$. This angle--average is shown in appendix A, together with the expressions
that define the Pauli operator in a particular $(T,M_T)$
channel in terms of the basis of physical states. Although we keep the index $M_T$ in
the bare
potential matrix elements they do not really have a dependence on
the third component
of isospin since we consider charge symmetric and
charge independent interactions. Therefore, the dependence of the
$G$--matrix on the third component of isospin comes exclusively
from 
the Pauli operator, since, as can be clearly seen in appendix A,
it acquires a dependence on $M_T$ when
different concentrations 
of particles belonging to the same isomultiplet 
(i.e. different values for the
corresponding $k_F$'s) are considered.

In comparison with the pure nucleonic calculation, this problem
is
a little bit more complicated because of its coupled--channel
structure.
Whereas for the strangeness sectors $0$ and $-4$ there is only one particle
channel
($NN$ and $\Xi$$\Xi$ respectively) and two
possible isospin states ($T=0,1$), in the $S=-1$($S=-3$) sector we are
dealing
with the $\Lambda N$($\Lambda \Xi$) and $\Sigma N$($\Sigma \Xi$) channels, coupled
to
$T=1/2$

\[\left(\begin{array}{cc}
G_{{\Lambda}N\rightarrow{\Lambda}N} & 
G_{{\Lambda}N\rightarrow{\Sigma}N} \\
G_{{\Sigma}N\rightarrow{\Lambda}N} & 
G_{{\Sigma}N\rightarrow{\Sigma}N}        
        \end{array}
  \right) \, \, \, \, \,
\left(\begin{array}{cc}
G_{{\Lambda}{\Xi}\rightarrow{\Lambda}{\Xi}} & 
G_{{\Lambda}{\Xi}\rightarrow{\Sigma}{\Xi}} \\
G_{{\Sigma}{\Xi}\rightarrow{\Lambda}{\Xi}} & 
G_{{\Sigma}{\Xi}\rightarrow{\Sigma}{\Xi}}        
        \end{array}
  \right) \ ,\] \, 
and the $\Sigma N$($\Sigma \Xi$) channel in isospin $T=3/2$

\[\left(\begin{array}{cc}
G_{{\Sigma}N\rightarrow{\Sigma}N}        
        \end{array}
  \right) \, \, \, \, \,
\left(\begin{array}{cc}
G_{{\Sigma}{\Xi}\rightarrow{\Sigma}{\Xi}}        
        \end{array}
  \right) \ .\] \,
In the $S=-2$ sector we must consider the channels
$\Lambda$$\Lambda$, $\Lambda$$\Sigma$, $\Xi$$N$ and
$\Sigma$$\Sigma$ in 
isospin states $T=0$ 

\[\left(\begin{array}{ccc}
G_{{\Lambda}{\Lambda}\rightarrow{\Lambda}{\Lambda}} &
G_{{\Lambda}{\Lambda}\rightarrow{\Xi}N} &
G_{{\Lambda}{\Lambda}\rightarrow{\Sigma}{\Sigma}}  \\
G_{{\Xi}N\rightarrow{\Lambda}{\Lambda}} &
G_{{\Xi}N\rightarrow{\Xi}N} &
G_{{\Xi}N\rightarrow{\Sigma}{\Sigma}}  \\
G_{{\Sigma}{\Sigma}\rightarrow{\Lambda}{\Lambda}} &
G_{{\Sigma}{\Sigma}\rightarrow{\Xi}N} &
G_{{\Sigma}{\Sigma}\rightarrow{\Sigma}{\Sigma}}  
        \end{array}
  \right) \ ,\] 
$T=1$

\[\left(\begin{array}{ccc}
G_{{\Xi}N\rightarrow{\Xi}N} &
G_{{\Xi}N\rightarrow{\Lambda}{\Sigma}} &
G_{{\Xi}N\rightarrow{\Sigma}{\Sigma}}  \\
G_{{\Lambda}{\Sigma}\rightarrow{\Xi}N} &
G_{{\Lambda}{\Sigma}\rightarrow{\Lambda}{\Sigma}} &
G_{{\Lambda}{\Sigma}\rightarrow{\Sigma}{\Sigma}}  \\
G_{{\Sigma}{\Sigma}\rightarrow{\Xi}N} &
G_{{\Sigma}{\Sigma}\rightarrow{\Lambda}{\Sigma}} &
G_{{\Sigma}{\Sigma}\rightarrow{\Sigma}{\Sigma}}  
        \end{array}
  \right) \ ,\] 
and $T=2$

\[\left(\begin{array}{cc}
G_{\Sigma\Sigma\rightarrow\Sigma\Sigma}        
        \end{array}
  \right)\ .\]  

In addition, each box $G_{{B_1B_2}\rightarrow{B_3B_4}}$ has 
a $2\times2$ 
matrix sub-structure to incorporate the couplings between 
$(L,S)$ states
having the same total angular momentum $J$. 
This sub-matrix reads

\[\left(\begin{array}{cc}
\left\langle L=J,S=0\right |
      G
\left | L=J,S=0 \right\rangle &
\left\langle L=J,S=0\right |
      G
\left | L=J,S=1 \right\rangle \\
\left\langle L=J,S=1\right |
      G
\left | L=J,S=0 \right\rangle &
\left\langle L=J,S=1\right |
      G
\left | L=J,S=1 \right\rangle   
        \end{array}
  \right) \ ,\]  
for spin singlet--spin triplet coupling  ($L=J,S=0 \leftrightarrow
L=J,S=1$) and 
\[\left(\begin{array}{cc}
\left\langle L=J-1,S=1\right |
      G
\left | L=J-1,S=1 \right\rangle &
\left\langle L=J-1,S=1\right |
      G
\left | L=J+1,S=1 \right\rangle \\
\left\langle L=J+1,S=1\right |
      G
\left | L=J-1,S=1 \right\rangle &
\left\langle L=J+1,S=1\right |
      G
\left | L=J+1,S=1 \right\rangle   
        \end{array}
  \right)\]  
for tensor coupling ($L=J-1,S=1 \leftrightarrow L=J+1,S=1$).

\subsection{The baryon single--particle energy in
Brueckner--Hartree--Fock
approximation} 
In the Brueckner--Hartree--Fock approximation the single--particle
potential of a
baryon $B_1$ which is embedded in the Fermi sea of baryons $B_2$
is given,
using the partial wave decomposition of the $G$--matrix, by
\begin{eqnarray}
 & & U_{B_1}^{(B_2)}(k_{B_1})= \nonumber \\
& & \frac{(1+\xi_{B_1})^3}{2}\sum_{J,l,S,T,M_T}
(2J+1)(1-(-1)^{L+S+T-S_{B_1}-S_{B_2}-T_{B_1}-T_{B_2}}) \nonumber \\
& & \times \mid \langle T_{B_1}T_{B_2}M_T^{(B_1)}M_T^{(B_2)} |
TM_T \rangle \mid^2
\int_{0}^{k_{max}}k^2dk f(k,k_{B_1}) \nonumber \\
& & \times  \left\langle B_1 B_2; kKLSTM_T\right | G^{J}
(E_{B_1}(k_{B_1})+E_{B_2}(k_{B_2})+m_{B_1}+m_{B_2})\left |
B_1 B_2; kKLSTM_T\right\rangle  \ ,
\label{eq:upot1}
\end{eqnarray}
if both types of baryons are identical, or by
\begin{eqnarray}
 & &  U_{B_1}^{(B_2)}(k_{B_1})= \nonumber \\
& & \frac{(1+\xi_{B_1})^3}{2}\sum_{J,l,S,T,M_T}
(2J+1)\mid\langle T_{B_1}T_{B_2}M_T^{(B_1)}M_T^{(B_2)} | TM_T
\rangle\mid^2
\int_{0}^{k_{max}}k^2dk f(k,k_{B_1})
\nonumber \\
& & \times   \left\langle B_1 B_2; kKLSTM_T\right | G^{J}
(E_{B_1}(k_{B_1})+E_{B_2}(k_{B_2})+m_{B_1}+m_{B_2})\left |
B_1 B_2; kKLSTM_T\right\rangle  \ ,
\label{eq:upot2}
\end{eqnarray}
if they are different. The labels
$S_{B_1}, S_{B_2}$ ($T_{B_1}, T_{B_2}$) denote the spin (isospin)
of baryons $B_1$ and $B_2$,
respectively, and
$\langle T_{B_1}T_{B_2}M_T^{(B_1)}M_T^{(B_2)} |
T M_T \rangle$ is the Clebsch--Gordan coefficient coupling to 
total isospin $T$. The variable $k$ denotes the relative momentum
of the
$B_1$$B_2$
pair, which is constrained by 
\begin{equation}
   k_{max} = \frac{k_{F}^{(B_2)}+\xi_{B_1}k_{B_1}}{1+\xi_{B_1}} \ ,
\nonumber
\end{equation}
with $\xi_{B_1}=M_{B_2}/M_{B_1}$.
Finally, the weight function $f(k,k_{B_1})$, given by 
\begin{equation}
f(k,k_{B_1})= \left\{ \begin{array}{cl} 1 & {\rm for\ }  k\leq
\frac{k_{F}^{(B_2)}-\xi_{B_1}k_{B_1}}{1+\xi_{B_1}} , \\ 0 &
{\rm for\ } 
|\xi_{B_1}k_{B_1}-(1+\xi_{B_1})k| > k_{F}^{(B_2)} , \\
\displaystyle\frac{{k_{F}^{(B_2)}}^2-[\xi_{B_1}k_{B_1}-(1+\xi_{B_1})k]^2}{4\xi
_{B_1}(1+\xi_{B_1})k_{B_1}k
} & \mbox{otherwise,}

\end{array} \right. \nonumber
\end{equation}
results from the analytical angular integration, once the angular dependence of the $G$-matrix
elements is eliminated.
This is done by choosing appropriate angular averages for
the center-of-mass of the $B_1B_2$ pair and for
the value of $k_{B_2}$ which enters
in the determination of the starting energy. See appendix B for details. 

If the baryon $B_i$ is embedded 
in the Fermi seas of several baryons $B_1$, $B_2$,
$B_3$, $\dots$, including its own Fermi sea, then its
single--particle potential is 
given by the sum of all the partial contributions
\begin{equation}
U_{B_i}(k)=\sum_{B_j}U_{B_i}^{(B_j)}(k)
\end{equation}
where $U_{B_i}^{(B_j)}(k)$ 
is the potential of the baryon $B_i$ due to the Fermi sea of
baryons $B_j$. In
this expression $k$ denotes the single--particle momentum of
particle $B_i$. The 
non--relativistic single--particle energy of baryon $B$ is then
given by
\begin{equation}
E_{B}(k)
=\frac{\hbar^2k^2}{2M_{B}}+U_{B}(k)
\label{eq:spenergy}
\end{equation}
This is precisely the single--particle energy that determines
the value of the starting energy $\omega$ at which the 
$G_{B_1B_2{\leftrightarrow}B_3B_4}$-matrix in Eq.
(\ref{eq:upot1}) (or
(\ref{eq:upot2}))
should be
evaluated. This implies a self-consistent solution of
Eqs. (\ref{eq:gmat}), (\ref{eq:upot1}) (or (\ref{eq:upot2})) and
(\ref{eq:spenergy}).
The Fermi energy of each species is determined by setting $k$ to
the corresponding Fermi momentum in the above expression.

\subsection{Energy density and binding energy per baryon}
The total non--relativistic energy density, $\varepsilon$, and
the total binding
energy per baryon, $B/A$, can be evaluated from the baryon
single--particle
potentials in the following way
\begin{equation}
\varepsilon=2\sum_{B}
\int_0^{k_F^{(B)}} \frac{d^3 k}{(2\pi)^3} 
\left( \frac{\hbar^2k^2}{2M_B}+\frac{1}{2}U_B(k) \right)
\label{eq:binding}
\end{equation}
\begin{equation}
\frac{B}{A}=\frac{\varepsilon}{\rho} \ ,
\label{eq:binding2} 
\end{equation}
where $\rho$ is the total barionic density. The density of a given baryon species is given by
\begin{equation}
\rho_{B}=\frac{k_{F_B}^3}{3\pi^2}=x_{B}\rho \ ,
\end{equation}
where $x_{B}=\rho_B/\rho$ is the fraction of
baryon
$B$, which is of course constrained by
\begin{equation}
\sum_{B}x_{B}=1 \ .
\end{equation}

\section{Results}
\label{sec:results}

We start this section by presenting results for the single--particle
potential of each baryon species, as a function of the baryon
momentum, for several baryonic densities and various nucleonic
and
hyperonic fractions. We have focussed on results for the Nijmegen
model (e) of the recent parametrization \cite{sr99}, since it
gives, together with model (f), the best predictions for
hypernuclear observables \cite{rsy98}, apart
from reproducing the $YN$ scattering scattering data as
well as the other models. We will restrict our calculations to matter composed of
neutrons, protons, $\Lambda$'s and $\Sigma^-$'s, since these last two hyperons
species are the first ones to appear \cite{schulze3}. This is also
confirmed on the recent study of $\beta$--stable neutron star matter
\cite{isaac99} up tp baryonic density $1.2$ fm$^{-3}$.

\begin{figure}
       \setlength{\unitlength}{1mm}
       \begin{picture}(100,180)
       \put(15,10){\epsfxsize=12cm \epsfbox{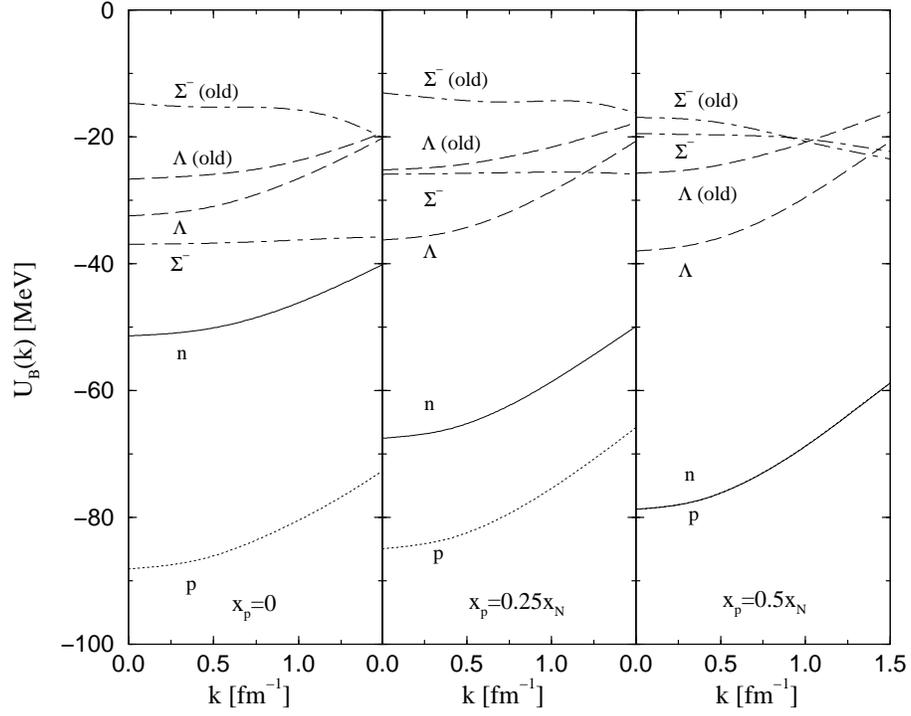}}
       \end{picture}
\caption{Momentum dependence of the single--particle potentials 
for the different species at $\rho=0.17$ fm$^{-3}$, hyperon fraction $x_Y=0$
and several
nucleon asymmetries.}
\label{fig:ukrho0}
\end{figure}

In Fig. \ref{fig:ukrho0} we show our results for 
non-strange nuclear matter at normal density, $\rho_0=0.17$
fm$^{-
3}$, and three proton fractions ($x_p=0.5x_N$, $0.25x_N$ and
0),
where $x_N$ is the fraction of non-strange baryons, which in this
case is 1. We also show the hyperon single--particle potentials,
denoted with the label ``old", obtained with the Nijmegen 1989
version of the $YN$ interaction \cite{nijme89}. On the right
panel, corresponding to symmetric
nuclear matter, we see that neutrons and protons have the same
single--particle potential, of the order of $-79$ MeV at zero
momentum.
Looking at the middle and left panels we see how, as the
fraction of protons decreases, the protons gain binding while the
neutrons lose attraction. This is a consequence of the different
behavior of the $NN$ interaction in the $T=0$ and $T=1$ channels,
the $T=0$ channel being substantially more
attractive. The potential of the proton is built from more
$T=0$
than $T=1$ pairs and hence becomes more attractive. 
The $\Lambda$ single--particle potential in
symmetric nuclear matter turns out to be around $-38$ MeV at
$k=0$ and has a smooth parabolic behavior as a function of $k$.
This result is larger than the value of $-30$ MeV obtained
when one extrapolates to large $A$  the s-wave $\Lambda$ 
single--particle energy of several hypernuclei \cite{bmz90}. It is also
much larger in magnitude than the value of around $-24$ MeV
\cite{morten96,isa98,yama94} which is obtained using the 1989
version of the Nijmegen $YN$ potential \cite{nijme89} with the
standard choice for the spectrum of the intermediate $YN$ states
in the Bethe-Goldstone equation. 
The value of the $\Sigma¯$ single--particle potential at
$k=0$ of $-20$
MeV is somewhat more attractive than that obtained with the 1989
potential of around $-17$ MeV. The function
$U_{\Sigma^-}(k)$ remains pretty constant in the range of momenta
explored.
Apart from the different size, the new single--particle hyperon
potentials also show a totally different behavior with increasing
asymmetry than that observed for the potentials obtained  with
the 1989 Nijmegen $YN$ interaction.  While the old $\Lambda$
single--particle potential turns to be slightly more attractive
with increasing neutron fraction (i.e. going from the right panel
to the left one), the new one becomes slightly more repulsive.
The changes for the $\Sigma^-$ single--particle potential are more
drastic. While the 1989 interaction gives a $\Sigma^-$ potential which
shows a little change with increasing neutron fraction, the
new $\Sigma^-$ potential becomes strongly attractive. The value
at $k=0$ for the $\Sigma^-$ potential changes from about $-20$
MeV in symmetric nuclear matter to $-37$ MeV in neutron matter.
This has important consequences in the composition of dense
matter: if 
hyperons feel substantially more attraction, their appearance in
dense
matter will happen at lower density. We note that our results
with the 1989 Nijmegen interaction are consistent with those
shown in \cite{schulze3}, where the same $YN$ interaction is
used.
Some differences are
found in the magnitude of the single--particle potentials which
should be adscribed to the use of a continuum spectrum
prescription in the case of \cite{schulze3}.

\begin{figure}
       \setlength{\unitlength}{1mm}
       \begin{picture}(100,180)
       \put(15,10){\epsfxsize=12cm \epsfbox{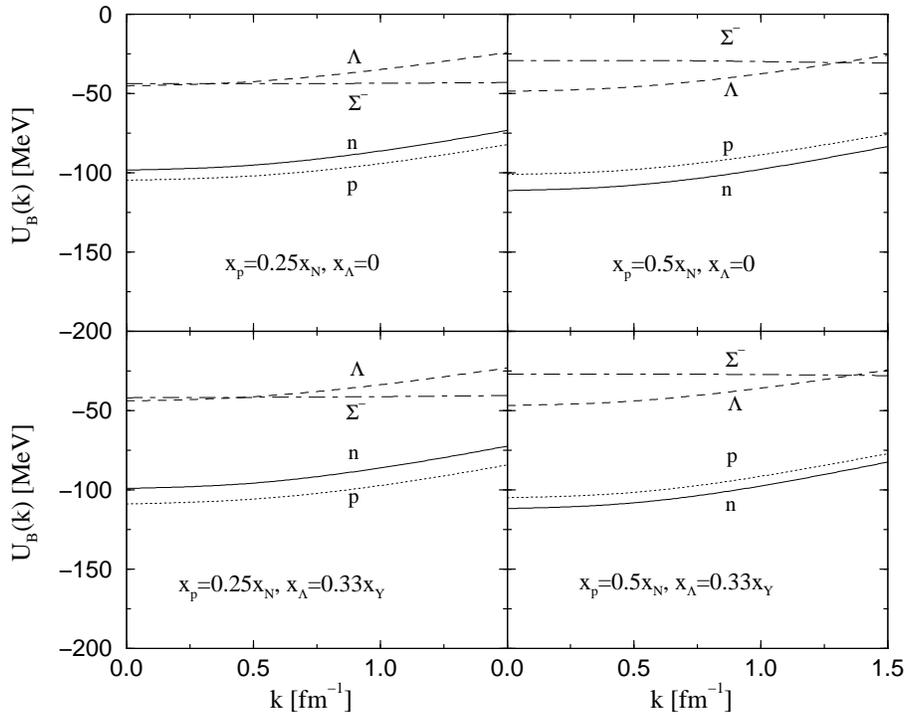}}
       \end{picture}
\caption{Momentum dependence of the single--particle potentials
for the different
species at $\rho=0.3$ fm$^{-3}$ and hyperon fraction $x_Y=0.1$.
The right panels 
correspond to symmetric nuclear matter, $x_n=x_p=0.5 x_N$, while
the left ones
are for asymmetric nuclear matter with $x_n=3x_p=0.75 x_N$. In the top 
panels the hyperonic fraction is built exclusively from
$\Sigma^-$ ($x_{\Sigma^-}=x_Y$) while in the bottom ones
there is a fraction of $\Lambda$'s
($x_{\Lambda}=x_{Y}/3$) and $\Sigma^{-}$'s ($x_{\Sigma^-}=2x_Y/3$)}
\label{fig:ukrho03}
\end{figure}

\begin{figure}
       \setlength{\unitlength}{1mm}
       \begin{picture}(100,180)
       \put(15,10){\epsfxsize=12cm \epsfbox{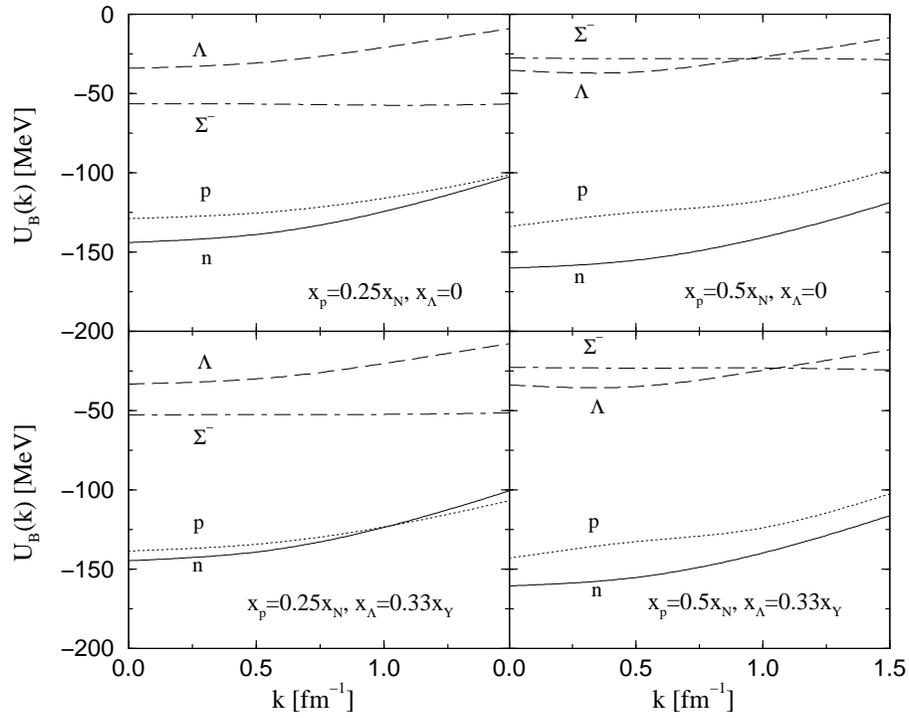}}
       \end{picture}
\caption{
The same as Fig.~\protect\ref{fig:ukrho03} for a baryon
density
$\rho=0.6$ fm$^{-3}$}
\label{fig:ukrho06}
\end{figure}

Having established how the nucleons affect the single--particle
potential of hyperons it is necessary to investigate the
influence of a finite fraction of hyperons on
the hyperons themselves and on the nucleons.
This is visualized in Figs. \ref{fig:ukrho03}
and \ref{fig:ukrho06} that show the single--particle potentials of
the different baryons as functions of the momentum. Figure
\ref{fig:ukrho03} shows results at
$\rho=0.3$ fm$^{-3}$ and a hyperon fraction $x_Y=0.1$, which is
assumed to come from only $\Sigma^-$ (top panels) or split into
$\Sigma^-$ and $\Lambda$ hyperons in a proportion $2:1$, hence
$x_{\Sigma^-}=2x_Y/3$ and $x_\Lambda=x_Y/3$ (bottom panels). The
panels on the right correspond to symmetric
proton-neutron composition ($x_p=x_n=0.5x_N$, where $x_N=0.9$)
and the ones on the left correspond to a higher proportion of
neutrons ($x_p=0.25x_N$, $x_n=0.75x_N$).
Starting at the upper-right panel we observe that the presence
of $\Sigma^-$ hyperons already breaks the symmetry between the
proton and neutron single--particle potentials in a symmetric
nucleonic composition, the neutrons
feeling around $-10$ MeV more attraction. This is due to a
different behavior between the $\Sigma^- n$ interaction which
only happens via the attractive $T=3/2$ channel 
and the $\Sigma^-p$ interaction that also receives contributions
from the very repulsive $T=1/2$ $\Sigma N$ component. In fact,
the difference bewteen the neutron and proton potentials is not
as pronounced as we move to the lower panel on the right, where
some $\Sigma^-$ hyperons are replaced by $\Lambda$ hyperons which
act identically over protons and neutrons. In the upper left
panel, where we have increased the neutron fraction in the
non-strange sector, we observe the typical pattern for
the nucleon single--particle potentials commented in Fig.
\ref{fig:ukrho0}: the particle with the smallest fraction (i.e. the proton)
shows
more binding. However, this
behavior is partially compensated by the presence of a sea of
$\Sigma^-$ which provides attraction (repulsion) to the neutron
(proton) single--particle potential. 
We also observe that the $\Sigma^-$ feels more
attraction, as a consequence of having replaced some repulsive
$\Sigma^- p$ pairs by attractive $\Sigma^- n$ ones. The $\Lambda$
loses binding because the Fermi sea of neutrons is larger and
their contribution to the $\Lambda$ single--particle energy
explores higher relative momentum components of the effective
$\Lambda n$ interaction, which are less attractive than the small
relative momentum ones.
Finally,
since
the Fermi sea of hyperons is small, the differences observed on
the potentials by going from the top panels to the
corresponding lower ones (which amounts to replacing $\Sigma^-$
hyperons by $\Lambda$ ones) are also small. 

Similar effects are found in the results reported in Fig.
\ref{fig:ukrho06}, which have been obtained for a baryonic
density $\rho=0.6$ fm$^{-3}$, where it is 
expected that nuclear matter in $\beta$ equilibrium
already contains hyperons \cite{schulze3}. 
The single--particle potential of the $\Lambda$
hyperon is less attractive than that for $\rho=0.3$ fm$^{-3}$ while that
of the $\Sigma^-$ is very similar.
It just gains somewhat more attraction when
the number of neutrons increase relative to that of protons in
going from the right panels to the left ones. As for the nucleon
single--particle potentials we observe, also on the left panels,
that the attractive $\Sigma^- n$ interaction is enhanced at
these high densities and makes the neutron spectrum more
attractive than the proton one, even in the asymmetric situation
when one would expect the protons to be more bound.

\begin{figure}
       \setlength{\unitlength}{1mm}
       \begin{picture}(100,180)
       \put(15,10){\epsfxsize=12cm \epsfbox{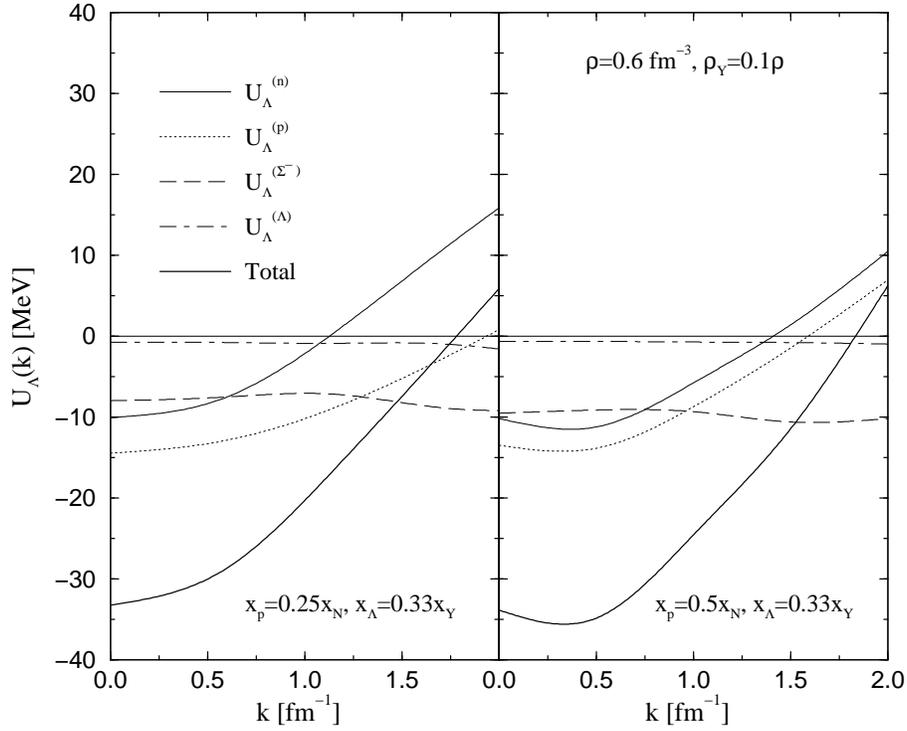}}
       \end{picture}
\caption{Separate contributions of each species to the
$\Lambda$ single--particle potential at  
$\rho=0.6$ fm$^{-3}$ and hyperon fraction $x_Y=0.1$ split
into $x_{\Sigma^-}=2 x_Y/3$ and $x_{\Lambda}=x_Y/3$. The right
panel is for
symmetric nuclear matter ($x_n=x_p=0.5 x_N$) and the left one 
for asymmetric nuclear matter 
($x_n= 3 x_p=0.75 x_N$).} 
\label{fig:ulamcon}
\end{figure}

\begin{figure}
       \setlength{\unitlength}{1mm}
       \begin{picture}(100,180)
       \put(15,10){\epsfxsize=12cm \epsfbox{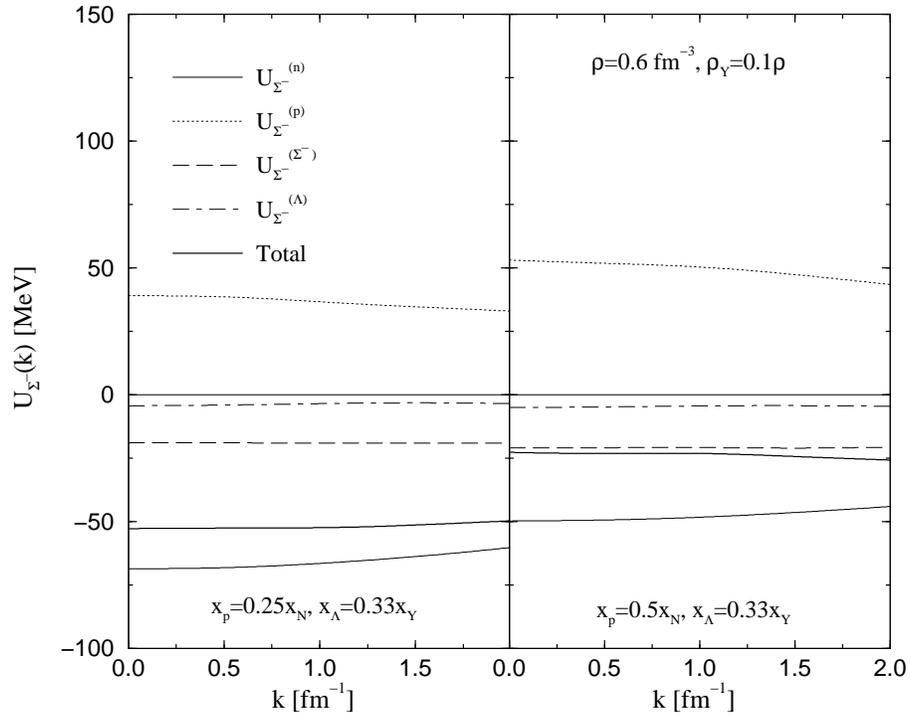}}
       \end{picture}
\caption{
The same as Fig~\protect\ref{fig:ulamcon} for the
$\Sigma^-$ single--particle potential.}
\label{fig:usigcon}
\end{figure}

To assess the influence of the $YY$ interaction we represent the
separate contributions building the $\Lambda$ single--particle
potential
in Fig. \ref{fig:ulamcon} and those for the $\Sigma^-$ one in
Fig.
\ref{fig:usigcon}, for a baryonic density of $0.6$ fm$^{-3}$.
The hyperon fraction of $x_Y=0.1$ is split
into fractions $x_{\Sigma^-}=2x_Y/3$ and 
$x_\Lambda=x_Y/3$ for 
$\Sigma^-$ and $\Lambda$ hyperons, respectively.
The results on the right hand side of Figs. \ref{fig:ulamcon} and
\ref{fig:usigcon}
correspond to the symmetric nuclear case and those on the left to
a neutron fraction three times larger than that of protons. We
see that the
contribution to the $\Lambda$ potential from the $\Lambda$
hyperons, represented by the dash-dotted line, is attractive and
almost negligible, due to a weak attractive $\Lambda\Lambda$
effective interaction \cite{sl99} and to the small amount of
$\Lambda$ particles present. On the contrary, the contribution
from the $\Sigma^-$ hyperons is larger,
of the order of $-10$ MeV in nuclear-symmetric matter and
slightly less in nuclear-asymmetric one,
which is comparable in size with the contribution from protons
and neutrons. This example clearly shows the important role of
the $YY$
interaction in modifying the properties of the $\Lambda$ hyperon.
The $\Lambda$ acquires more attraction and its appearance in
dense matter becomes more favorable with respect to the situation
in which the $YY$ interaction was neglected.
The fact that the neutron (thin solid line) and proton (dotted
line) contributions to the $\Lambda$ single--particle potential
are not the same in nuclear-symmetric matter is due to the
$\Sigma^-$ hyperons which make
the neutrons feel more
attraction and, consequently, the $\Lambda n$ pairs explore the
effective
$\Lambda N$ interaction at smaller energies, where it is less
attractive.
The different contributions to the $\Sigma^-$ potential
are shown in Fig. \ref{fig:usigcon}. The 
$\Lambda$ hyperons (dot-dashed line) contribute very little
due to the reduced value of their Fermi momentum.
The contribution of the
$\Sigma^-\Sigma^-$ pairs (long-dashed line) is very important, of
the order of $-25$ MeV in symmetric nuclear matter, and becomes
crucial due to the fact that the neutron (thin solid line) and
the proton (dotted line) contributions, which amount each one to
about 50 MeV in magnitude, almost cancel each other. In the left
panel, the replacement of protons by neutrons, lowers the
$\Sigma^-$ single--particle potential considerably, by about $25$
MeV.  Again, neglecting
the $YY$ interactions here would have made the $\Sigma^-$
potential about $20-25$ MeV less attractive, retarding its
appearance
in dense matter.

\begin{figure}
       \setlength{\unitlength}{1mm}
       \begin{picture}(100,180)
       \put(15,10){\epsfxsize=12cm \epsfbox{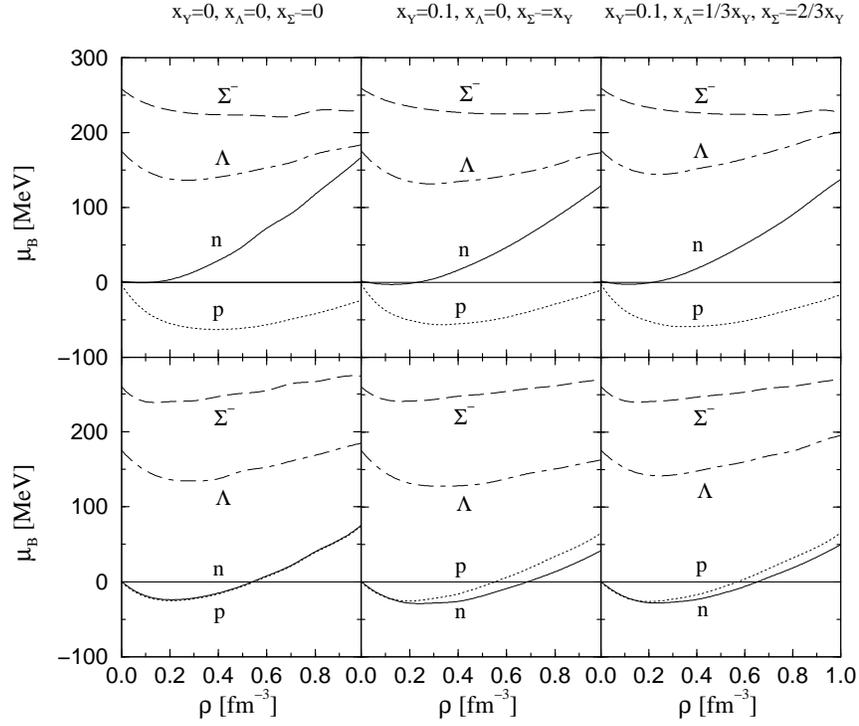}}
       \end{picture}
\caption{Chemical potentials of the different species as 
functions of 
total baryonic density, for different nucleonic asymmetries and 
strangeness fractions. The top panels correspond to the asymmetric nuclear matter case
($x_n=3 x_p=0.75x_N$), while the bottom ones correspond to symmetric nuclear matter
($x_n=x_p=0.5x_N$).}
\label{fig:chemical}
\end{figure}

The analysis of the structure of 
$\beta$-stable matter requires the knowledge of the
chemical potential ($\mu_B$) of each baryon,
defined at zero temperature as the 
single--particle energy of the Fermi momentum 
[Eq. (\ref{eq:spenergy})]. 
In Fig. \ref{fig:chemical} we  show the
chemical potentials as functions of density for different nucleon
asymmetries and
hyperon fractions. Note that the curves
are measured with respect to the nucleon mass and contain, in
addition to
the non-relativistic Fermi energy,
the baryon mass of each species. 
The top panels show the results for asymmetric nuclear matter
($x_n=3 x_p= 0.75 x_N$)
whereas the bottom panels stand for the symmetric case.  
On the left panels we show 
results for purely nucleonic matter ($x_Y=0$), on the central
panels we have 
$x_{\Sigma^-}=x_Y=0.1$, while on the right panels $x_Y$ is
distributed into
$x_{\Lambda}=x_Y/3$ and $x_{\Sigma^-}=2x_Y/3$. 
The behavior of the chemical potentials when increasing the
nucleonic asymmetry
as well as the hyperonic fraction follows closely the trends
observed in 
Figs.~\ref{fig:ukrho0}, \ref{fig:ukrho03} and \ref{fig:ukrho06}
for the single--particle potential at densities $\rho=0.17$, 0.3
and
0.6 fm$^{-3}$, respectively. We just have to consider here that
the curves in Fig.~\ref{fig:chemical} also contain
the kinetic
energy of the corresponding Fermi momentum.
It is interesting to comment on the high density behavior of the
chemical potentials, since this will determine the feasibility of
having hyperons in beta-stable neutron star matter.
In symmetric nuclear matter, both the $\Lambda$ and the
$\Sigma^-$
chemical potentials show, from a certain density on,
an increase with increasing density which is very mild as
compared to that assumed by phenomenological
$YN$ interactions \cite{MDG}.
When the number
of neutrons over that of protons is increased (top panels), the
$\Lambda$
chemical potential barely changes because of the similarity
between the $\Lambda n$
and $\Lambda p$ interaction. However, the $\Sigma^-$ hyperon
acquires 
more binding due to the dominant $\Sigma^- n$ attractive pairs
over
the $\Sigma^- p$ repulsive ones.
This will favor the appearance of $\Sigma^-$ in dense neutron
star matter,
through the $n n \rightarrow p \Sigma^-$ conversion, when the
equilibrium between chemical chemical
potentials is achieved at both sides. Once a Fermi sea of $\Sigma^-$
hyperons
starts to build up, however, 
the neutrons become more attractive moderating, in turn, the
appearance
of $\Sigma^-$ hyperons. 
As we see, the composition of dense neutron star matter in
equilibrium will
result from a delicate interplay between the mutual influence
among the different species. In fact,
one needs to find, at each baryonic density, the particle
fractions which balance the chemical potentials
in the weak and strong reactions that transform the
species among themselves. This study, which is beyond the scope
of the present work, will be presented in a separate publication
\cite{isaac99}.

One of the novelties of this work is that
we allow for different concentrations for the baryon species.
Therefore, we can
explicitly treat the dependence of the $G$-matrix on the third
component of isospin which comes from the Pauli operator of 
species $B,\widetilde{B}$ that may have, even when belonging to 
the same 
isospin-strangeness multiplet, different Fermi momenta. 
See appendix A for more details. 

In Fig. \ref{fig:gmatrix} we report 
the diagonal $\Sigma N \to \Sigma N$ $G$-matrix elements
in the $^1S_0$ channel, as a functiom of relative momentum for a
density $\rho=0.6$ fm$^{-3}$, taking
 $x_{\Lambda}=0$ and $x_{\Sigma^-}=x_Y=0.1$. The top panels
correspond to 
the isospin $T=1/2$ channel and the lower ones to the $T=3/2$
one. The panels on the right 
are for symmetric nuclear matter, $x_n=x_p=0.5 x_N$, while those on
the left
correspond to $x_p=0.25 x_N$. The starting energy and center of
mass is the same for all the curves shown in the same plot, thus the
dependence on $M_T$ comes exclusively from the Pauli operator.
Note that different pairs of particles contribute to each
$(T,M_T)$ combination. The case $(T,M_T)=(1/2,+1/2)$ receives
contributions from
$\Sigma^+ n$ and $\Sigma^0 p$ pairs while $\Sigma^- p$
and $\Sigma^0 n$ contribute to 
$(T,M_T)=(1/2,-1/2)$. In the case of isospin $T=3/2$ one has
contributions from $\Sigma^- n$ ($M_T=-3/2$),
$\Sigma^0 n,\Sigma^- p$ ($M_T=-1/2$),
$\Sigma^+ n,\Sigma^0 p$ ($M_T=+1/2$) and
$\Sigma^+ p$ ($M_T=+3/2$).
We observe that the curve corresponding to the third component
$M_T$ less affected by Paui blocking is always more attractive as
the phase-space for intermediate states, which induce attractive
corrections to the potential matrix elements,
is larger. This is clearly seen in the top panel on the right,
since the dotted line contains a channel with the
$\Sigma^-$ hyperon.
When the
nucleonic asymmetry is increased by going to the panel on the
left, the effects of Pauli blocking on the neutrons are more
important than those on the $\Sigma^-$ hyperons. 
This is the reason for the full curve to appear above the
dotted one, since
the $(T,M_T)=(1/2,+1/2)$ case receives contributions from
$\Sigma^0 p$ and $\Sigma^+ n$ pairs in a proportion 1:2 and
it contains relatively more neutrons than the case 
$(T,M_T)=(1/2,-1/2)$ with 
$\Sigma^- p$ and $\Sigma^0 n$ pairs in a proportion 2:1.
In the case of $T=3/2$ we observe that the asymmetry on the
$\Sigma$ multiplet barely induces any dependence on $M_T$ in the
$G$-matrix, as can be seen from the bottom panel on right.
However, one can observe differences when going to asymmetric
nuclear matter on the left panel since the Pauli blocking on
$\Sigma^- n$ pairs ($M_T=-3/2$) is enhanced over that on
$\Sigma^+ p$ pairs ($M_T=+3/2$).
As we can see,
in all cases considered here the dependence of the $G$-matrix on the
third component of
the isospin is very weak and can almost be neglected. We
have also encountered this weak dependence in the other
$B\widetilde{B}$ $G$-matrices.  
Therefore, a presumably good strategy and less time consuming
would be to 
obtain the $G$-matrices in isospin saturated systems and,
afterwards, calculate the single--particle potentials by folding
the 
``approximate" effective interactions with the different baryon
Fermi seas. 
 
\begin{figure}
       \setlength{\unitlength}{1mm}
       \begin{picture}(100,180)
       \put(15,10){\epsfxsize=12cm \epsfbox{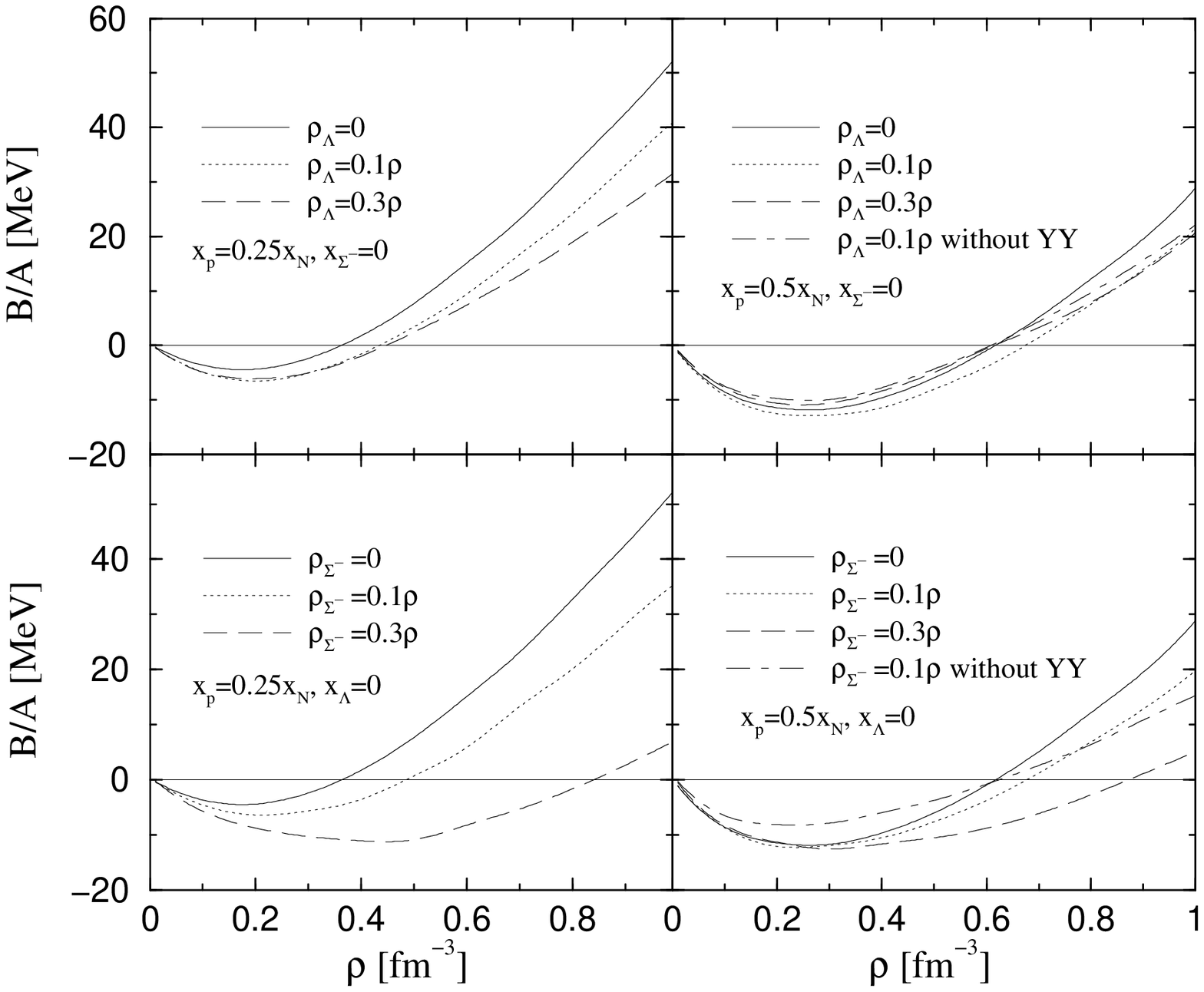}}
       \end{picture}
\caption{Binding energy per baryon as a function of the baryon
density. In the top panels we set
$x_{\Sigma^-}=0$ and show results for several values  
of $x_{\Lambda}$, while the bottom panels correspond
to $x_{\Lambda}=0$ and different fractions of $\Sigma^-$'s. 
The panels on the right 
are for symmetric nuclear matter, while the left ones correspond
to
asymmetric nuclear matter ($x_n=3 x_p=0.75 x_N$). In the case of nuclear
symmetric matter with 10 $\%$ of hyperons we also show a curve (dash-dotted line) where
the $YY$ 
interaction has been turned off.} 
\label{fig:binding}
\end{figure}

We finish this section by reporting in Fig. \ref{fig:binding} the
binding
 energy per baryon as a function of density. The right  and left
figures describe symmetric and asymmetric 
($x_n=3 x_p=0.75 x_N$) nuclear matter, respectively. In the top panels,
we show the
binding energy with $x_{\Sigma^-}=0$ for  several values of
$x_\Lambda$ while
in the bottom panels we consider $x_\Lambda=0$ and vary the
concentration 
of $\Sigma^-$ hyperons.  The binding energy per baryon,
calculated according to Eqs.
(\ref{eq:binding}),(\ref{eq:binding2}), is 
the result of a balance between the average kinetic energy 
of each baryon Fermi sea and the  
contribution from the mutual interactions, given by the average
of the single--particle potential of each species. In order to
identify the effects of the $YY$
interaction on the binding energy per baryon we have
also included a curve corresponding to a calculation with a
10$\%$ 
of hyperons (either $\Lambda$'s or $\Sigma^-$'s) where the $YY$
interaction is turned off (dash-dotted line).  
In both cases, turning the $YY$-interaction on results in a gain
of binding
energy which is larger in the case of $\Sigma^-$. The binding
energy per baryon 
shows a saturation density, i.e. a density for which the
thermodynamic pressure
is zero, which is too high 
when we consider the composition with only nucleons. The
location
of this saturation density is little affected when the percentage
of hyperons
 is increased. When a small amount of nucleons is substituted by
hyperons there is automatically a decrease of the kinetic energy
contribution
because the hyperons can be accomodated in lower momentum states
and in 
addition have a larger bare mass. The analysis of the influence of the
effective interaction on the binding energy must be made separately
for $\Lambda$'s and $\Sigma^-$'s. 
Although the effective $\Lambda N$ and
$\Lambda \Lambda$ interactions are clearly less attractive than the $NN$
one, the reduction of kinetic energy is clearly enough to
compensate for the loss of binding energy when a 10\% of
nucleons is substituted by $\Lambda$'s.
Notice, however, that we have 
to consider the $\Lambda \Lambda$ interaction in order to obtain 
this increase of binding with respect to 
the pure nucleonic case. At $x_\Lambda=30$\% the loss of kinetic
energy is not enough to compensate for the loss of attraction
from the effective interactions and less binding 
energy than the case with only nucleonic degrees of freedom is
obtained.  Looking at the lower panels for the $\Sigma^-$
hyperons we observe that the binding energy per baryon gains
more attraction as compared to the case for $\Lambda$'s.
This is due, essentially, to the larger loss of kinetic energy
due to the larger mass of the $\Sigma^-$.
In general, the replacement of nucleons by hyperons produces a
gain in binding energy and
a softening of the equation of state. 
The appearance of hyperons in 
beta--stable matter, the softening
of the equation of state and its implications on the properties
of neutron stars are deferred to a future study \cite{isaac99}.

\section{Conclusions}
\label{conclusions}

In this work we have developed the formalism for microscopic
Brueckner-type calculations of dense nuclear 
matter with
strangeness, allowing for any concentration of the different baryon 
species.

By relating the Pauli operator to
the different pairs of physical particles that
contribute to the particular $(T,M_T)$ channel (see appendix A), we
have been able to obtain the $M_T$ dependence of the
effective interaction ($G$-matrix) between any two species.

We have seen, however, that the dependence of the
$G$-matrix on the third component of isospin is weak enough
to allow, in future studies, for a simpler strategy consisting
of obtaining the effective interactions in isospin
saturated situations ($k_F^{(n)}=k_F^{(p)}$, 
$k_F^{(\Sigma^-)}= k_F^{(\Sigma^0)}=k_F^{(\Sigma^+)}$,  
$k_F^{(\Xi^-)}= k_F^{(\Xi^0)}$ ).
The various single--particle potentials
can then be obtained by folding the
approximate effective interactions with the  
Fermi seas of the different species.  

We have studied the dependence of the 
single--particle potentials on the nucleon and
hyperon asymmetries, focussing on situations that can
be relevant in future studies of
beta-stable neutron star matter with strangeness.
This is why, apart from 
neutrons and protons, we have only
considered the $\Lambda$ and $\Sigma^-$ hyperons, which are
the first ones expected to appear.
We have compared the symmetric nuclear matter
composition ($x_n=x_p=x_N$) with the asymmetric case containing
a large fraction of neutrons ($x_n=3 x_p=0.75 x_N$), for a
small, but relevant, hyperon fraction $x_Y=0.1$. This fraction 
may be fully composed by $\Sigma^-$ hyperons ($x_{\Sigma^-}=x_Y$)
or contain also a small proportion of $\Lambda$'s 
($x_{\Sigma^-}= 2 x_{\Lambda} = 2 x_Y/3$).
We find that the presence of hyperons, especially $\Sigma^-$,
modifies substantially the single--particle potentials of the
nucleons. The neutrons feel an increased 
attraction due to the $\Sigma^- n$ effective interaction that only
happens through the very attractive $T=3/2$ $\Sigma N$ channel, 
while the protons feel a repulsion as the $\Sigma^- p$ pairs 
also receive contributions from the very repulsive
$T=1/2$ $\Sigma N$ one.

By decomposing the $\Lambda$ and $\Sigma^-$ single--particle potentials
in the contributions from the various species, we have 
seen the relevance of considering the $YY$ interaction.
For a baryonic density of $0.6$ fm$^{-3}$, a nuclear asymmetry
of $x_n=3 x_p=0.75 x_N$ and a hyperon
fraction of $x_Y=0.1$ (split into $x_{\Sigma^-}=2x_Y/3$ and
$x_\Lambda=x_Y/3$), we find that the hyperonic contribution to the
$\Lambda$ single--particle potential  at zero momentum
is of the order of $-10$ MeV
(1/3 of the total $U_{\Lambda}(0)$),
and that for the $\Sigma^-$ is of the order of $-25$ MeV
(1/2 of the total $U_{\Sigma^-}(0)$).

In the absence
of hyperonic Fermi seas the $\Lambda$ and $\Sigma^-$ chemical 
potentials show a mild increase with increasing baryonic density.
The presence of a Fermi sea of $\Sigma^-$ hyperons slows down 
this increase, especially for the $\Sigma^-$ chemical potential
and in the case of asymmetric nuclear matter, due
to the very attractive $T=3/2$ $\Sigma N$ interaction acting
on $\Sigma^- n$ pairs. This will make the balance between
chemical potentials in the
strong $n n \to \Sigma^- p$ conversion easier and will favor
the appearance of $\Sigma^-$ at lower densities.

Finally, we have studied the modifications of the binding energy
per baryon in symmetric and asymmetric nuclear matter when some
nucleons are replaced either by $\Lambda$ or $\Sigma^-$ hyperons.
As expected, we observe an increase 
in the binding energy, which increases with density, mainly as a result of 
a decrease in kinetic energy because the hyperons 
can be accommodated in lower momentum states and have a larger mass.
This effect will produce
a softening in the equation of state
that will influence the behavior of dense matter and the structure of neutron stars.

\section*{Acknowledgements}
This work is partially supported by the DGICYT contract No. PB95--1249
(Spain) and by the Generalitat de Catalunya grant No. 1998SGR--11. One of
the authors (I.V.) wishes to acknowledge support from a doctoral fellowship
of the Ministerio de Educaci\'on y Cultura (Spain).

\newpage
\renewcommand{\theequation}{\Alph{section}.\arabic{equation}}
\setcounter{section}{1}
\section*{{\bf Appendix A: Pauli operator in the different
strangeness channels}}

In this appendix we show how the Pauli operator
$Q_{B\widetilde{B}}$, which prevents scattering into
occupied $B\widetilde{B}$ intermediate states, acquires
a dependence on the third component of isospin due to the
different
Fermi momenta of baryons $B$ and $\widetilde{B}$.
The Pauli operator reads
\begin{equation}
Q_{B\widetilde{B}}(\vec{k},\vec{K})= \left\{ \begin{array}{cl} 1
& \mbox{for $
| \alpha\vec{K}+\vec{k} | > k_{F}^{B}$
and 
$| \beta\vec{K}-\vec{k} | > k_{F}^{\widetilde{B}}
 $ }, \\ 0 & \mbox{otherwise}
\end{array} \right. \nonumber
\setcounter{equation}{1}
\end{equation}
where $\vec{k}$ and $\vec{K}$ are, respectively, 
the relative and total momenta of the
$B\widetilde{B}$ pair,
$\alpha=\displaystyle\frac{m_{B}}{m_{B}+m_{\widetilde{B}}}$ 
and
$\beta=\displaystyle\frac{m_{\widetilde{B}}}{m_{B}+m_{\widetilde{
B}}}$. 
In order to solve the Bethe-Goldstone equation in
partial wave representation [see Eq. (\ref{eq:gmat})] we need to 
perform an angle average of the Pauli operator, which reads

\begin{equation}
\overline{Q}_{B\widetilde{B}}(k,K)= \left\{ \begin{array}{cl}
\frac{1}{2}(\cos{\theta}_{B}+\cos{\theta}_{\widetilde{B}}) 
& \mbox{if $ \cos{\theta}_{B}+\cos{\theta}_{\widetilde{B}}>0 $ }, 
\\ 
0 & \mbox{if $ \cos{\theta}_{B}+\cos{\theta}_{\widetilde{B}}<0 $ }
\end{array} \right. \nonumber
\end{equation}
where

\begin{equation}
\cos{\theta}_{B}= \left\{ \begin{array}{cl}
1 & \mbox{if $ | \alpha\vec{K}-\vec{k} | > k_{F}^{(B)} $ }, 
\\ 
\displaystyle\frac{{\alpha}^{2}K^{2}+{k}^{2}-{k_{F}^{(B)}}^{2}}{2{\alpha}Kk} &
\mbox{otherwise}
\end{array} \right. \nonumber
\end{equation}
and 

\begin{equation}
\cos{\theta}_{\widetilde{B}}= \left\{ \begin{array}{cl}
1 & \mbox{if $ | \beta\vec{K}-\vec{k} | > k_{F}^{({\widetilde{B}})} $ }, 
\\ 
\displaystyle\frac{{\beta}^{2}K^{2}+{k}^{2}-{k_{F}^{({\widetilde{B}})}}^{2}}{2{\beta}Kk} &
\mbox{otherwise}
\end{array} \right. \nonumber
\end{equation}

Taking the following convention for the isospin states
representing
the particle basis 
\begin{equation}
\left|n\right\rangle=\left|1/2,-1/2\right\rangle ; \\ 
\left|p\right\rangle=\left|1/2,+1/2\right\rangle  \nonumber
\end{equation}
\begin{equation}
\left|{\Lambda}\right\rangle=\left|0,0\right\rangle \nonumber
\end{equation}
\begin{equation}
\left|{\Sigma^-}\right\rangle=\left|1,-1\right\rangle ; \\
\hspace{0.25cm}
\left|{\Sigma^0}\right\rangle=\left|1,0\right\rangle ; \\
\hspace{0.25cm}
\left|{\Sigma^+}\right\rangle=-\left|1,+1\right\rangle  
\nonumber
\end{equation}
\begin{equation}
\left|{\Xi^-}\right\rangle=-\left|1/2,-1/2\right\rangle ; \\ 
\hspace{0.25cm}
\left|{\Xi^0}\right\rangle=\left|1/2,+1/2\right\rangle 
\nonumber
\end{equation}
it is easy to obtain the Pauli operator in the coupled-isospin
basis, $Q_{B\widetilde{B}}(k,K;T,M_T)$, for each strangeness
sector.
Note that in the following expressions we have only retained the 
dependence on the isospin labels.

{\bf {A.1. Strangeness $0$}}  

\vspace{0.25cm}
\begin{equation}
Q_{NN}(T=0,M_T=0)=\frac{1}{2}\left( Q_{pn}+Q_{np} \right) \\
\nonumber
\end{equation}
\begin{equation}
Q_{NN}(T=1,M_T=-1)=Q_{nn}
\nonumber
\end{equation}
\begin{equation}
Q_{NN}(T=1,M_T=0)=\frac{1}{2}\left( Q_{pn}+Q_{np} \right) \\
\nonumber
\end{equation}
\begin{equation}
Q_{NN}(T=1,M_T=+1)=Q_{pp} \\
\nonumber
\end{equation}

\vspace{0.25cm}
{\bf {A.2. Strangeness $-1$}}

\vspace{0.25cm}
\begin{equation}
Q_{{\Lambda}N}\left(T=\frac{1}{2},M_T=-\frac{1}{2}\right)=Q_{{\Lambda}n}
\nonumber
\end{equation}
\begin{equation}
Q_{{\Lambda}N}\left(T=\frac{1}{2},M_T=+\frac{1}{2}\right)=Q_{{\Lambda}p}
\nonumber
\end{equation}
\begin{equation}
Q_{{\Sigma}N}\left(T=\frac{1}{2},M_T=-\frac{1}{2}\right)=\frac{1}
{3}
Q_{{\Sigma^0}n}+\frac{2}{3}Q_{{\Sigma^-}p}
\nonumber
\end{equation}
\begin{equation}
Q_{{\Sigma}N}\left(T=\frac{1}{2},M_T=+\frac{1}{2}\right)=\frac{2}
{3}
Q_{{\Sigma^+}n}+\frac{1}{3}Q_{{\Sigma^0}p}
\nonumber
\end{equation}
\begin{equation}
Q_{{\Sigma}N}\left(T=\frac{3}{2},M_T=-\frac{3}{2}\right)=Q_{{\Sigma^-}n}
\nonumber
\end{equation}
\begin{equation}
Q_{{\Sigma}N}\left(T=\frac{3}{2},M_T=-\frac{1}{2}\right)=\frac{2}
{3}
Q_{{\Sigma^0}n}+\frac{1}{3}Q_{{\Sigma^-}p}
\nonumber
\end{equation}
\begin{equation}
Q_{{\Sigma}N}\left(T=\frac{3}{2},M_T=+\frac{1}{2}\right)=\frac{1}
{3}
Q_{{\Sigma^+}n}+\frac{2}{3}Q_{{\Sigma^0}p}
\nonumber
\end{equation}
\begin{equation}
Q_{{\Sigma}N}\left(T=\frac{3}{2},M_T=+\frac{3}{2}\right)=Q_{{\Sigma^+}p}
\nonumber
\end{equation}

\vspace{0.25cm}
{\bf {A.3. Strangeness $-2$}}

\vspace{0.25cm}
\begin{equation}
Q_{{\Lambda}{\Lambda}}(T=0,M_T=0)=Q_{{\Lambda}{\Lambda}}
\nonumber
\end{equation} 
\begin{equation}
Q_{{\Xi}N}(T=0,M_T=0)=\frac{1}{2}\left( Q_{{\Xi^-}p}
+Q_{{\Xi^0}n} \right)
\nonumber
\end{equation} 
\begin{equation}
Q_{{\Sigma}{\Sigma}}(T=0,M_T=0)=\frac{1}{3}\left( 
Q_{{\Sigma^+}{\Sigma^-}}
+Q_{{\Sigma^0}{\Sigma^0}}+Q_{{\Sigma^-}{\Sigma^+}} \right)
\nonumber
\end{equation} 
\begin{equation}
Q_{{\Xi}N}(T=1,M_T=-1)=Q_{{\Xi^-}n}
\nonumber
\end{equation} 
\begin{equation}
Q_{{\Xi}N}(T=1,M_T=0)=\frac{1}{2}\left( Q_{{\Xi^-}p}
+Q_{{\Xi^0}n} \right)
\nonumber
\end{equation} 
\begin{equation}
Q_{{\Xi}N}(T=1,M_T=+1)=Q_{{\Xi^0}p}
\nonumber
\end{equation} 
\begin{equation}
Q_{{\Lambda}{\Sigma}}(T=1,M_T=-1)=Q_{{\Lambda}{\Sigma^-}}
\nonumber
\end{equation} 
\begin{equation}
Q_{{\Lambda}{\Sigma}}(T=1,M_T=0)=Q_{{\Lambda}{\Sigma^0}}
\nonumber
\end{equation} 
\begin{equation}
Q_{{\Lambda}{\Sigma}}(T=1,M_T=+1)=Q_{{\Lambda}{\Sigma^+}}
\nonumber
\end{equation} 
\begin{equation}
Q_{{\Sigma}{\Sigma}}(T=1,M_T=-1)=\frac{1}{2}\left( 
Q_{{\Sigma^0}{\Sigma^-}}
+Q_{{\Sigma^-}{\Sigma^0}} \right)
\nonumber
\end{equation}
\begin{equation}
Q_{{\Sigma}{\Sigma}}(T=1,M_T=0)=\frac{1}{2}\left( 
Q_{{\Sigma^+}{\Sigma^-}}
+Q_{{\Sigma^-}{\Sigma^+}} \right)
\nonumber
\end{equation}
\begin{equation}
Q_{{\Sigma}{\Sigma}}(T=1,M_T=+1)=\frac{1}{2}\left( 
Q_{{\Sigma^0}{\Sigma^+}}
+Q_{{\Sigma^+}{\Sigma^0}} \right)
\nonumber
\end{equation}
\begin{equation}
Q_{{\Sigma}{\Sigma}}(T=2,M_T=-2)=Q_{{\Sigma^-}{\Sigma^-}}
\nonumber
\end{equation}
\begin{equation}
Q_{{\Sigma}{\Sigma}}(T=2,M_T=-1)=\frac{1}{2}\left( 
Q_{{\Sigma^0}{\Sigma^-}}+Q_{{\Sigma^-}{\Sigma^0}} \right)
\nonumber
\end{equation}
\begin{equation}
Q_{{\Sigma}{\Sigma}}(T=2,M_T=0)=\frac{1}{6}Q_{{\Sigma^+}{\Sigma^-
}}
+\frac{2}{3}Q_{{\Sigma^0}{\Sigma^0}} 
+\frac{1}{6}Q_{{\Sigma^-}{\Sigma^+}} 
\nonumber
\end{equation}
\begin{equation}
Q_{{\Sigma}{\Sigma}}(T=2,M_T=+1)=\frac{1}{2}\left( 
Q_{{\Sigma^0}{\Sigma^+}}+Q_{{\Sigma^+}{\Sigma^0}} \right)
\nonumber
\end{equation}
\begin{equation}
Q_{{\Sigma}{\Sigma}}(T=2,M_T=+2)=Q_{{\Sigma^+}{\Sigma^+}}
\nonumber
\end{equation}

\vspace{0.25cm}
{\bf {A.4. Strangeness $-3$}}  

\vspace{0.25cm}
\begin{equation}
Q_{{\Lambda}{\Xi}}\left(T=\frac{1}{2},M_T=-\frac{1}{2}\right)=Q_{{\Lambda}{\Xi^-}}
\nonumber
\end{equation}
\begin{equation}
Q_{{\Lambda}{\Xi}}\left(T=\frac{1}{2},M_T=+\frac{1}{2}\right)=Q_{{\Lambda}{\Xi^0}}
\nonumber
\end{equation}
\begin{equation}
Q_{{\Sigma}{\Xi}}\left(T=\frac{1}{2},M_T=-\frac{1}{2}\right)=\frac{1}
{3}
Q_{{\Sigma^0}{\Xi^-}}+\frac{2}{3}Q_{{\Sigma^-}{\Xi^0}}
\nonumber
\end{equation}
\begin{equation}
Q_{{\Sigma}{\Xi}}\left(T=\frac{1}{2},M_T=+\frac{1}{2}\right)=\frac{2}
{3}
Q_{{\Sigma^+}{\Xi^-}}+\frac{1}{3}Q_{{\Sigma^0}{\Xi^0}}
\nonumber
\end{equation}
\begin{equation}
Q_{{\Sigma}{\Xi}}\left(T=\frac{3}{2},M_T=-\frac{3}{2}\right)=Q_{{\Sigma^-}{\Xi^-}}
\nonumber
\end{equation}
\begin{equation}
Q_{{\Sigma}{\Xi}}\left(T=\frac{3}{2},M_T=-\frac{1}{2}\right)=\frac{2}
{3}
Q_{{\Sigma^0}{\Xi^-}}+\frac{1}{3}Q_{{\Sigma^-}{\Xi^0}}
\nonumber
\end{equation}
\begin{equation}
Q_{{\Sigma}{\Xi}}\left(T=\frac{3}{2},M_T=+\frac{1}{2}\right)=\frac{1}
{3}
Q_{{\Sigma^+}{\Xi^-}}+\frac{2}{3}Q_{{\Sigma^0}{\Xi^0}}
\nonumber
\end{equation}
\begin{equation}
Q_{{\Sigma}{\Xi}}\left(T=\frac{3}{2},M_T=+\frac{3}{2}\right)=Q_{{\Sigma^+}{\Xi^0}}
\nonumber
\end{equation}

\vspace{0.25cm}
{\bf {A.5. Strangeness $-4$}}  

\vspace{0.25cm}
\begin{equation}
Q_{{\Xi}{\Xi}}(T=0,M_T=0)=\frac{1}{2}\left( Q_{{\Xi^0}{\Xi^-}}+Q_{{\Xi^-}{\Xi^0}}
\right) \\
\nonumber
\end{equation}
\begin{equation}
Q_{{\Xi}{\Xi}}(T=1,M_T=-1)=Q_{{\Xi^-}{\Xi^-}}
\nonumber
\end{equation}
\begin{equation}
Q_{{\Xi}{\Xi}}(T=1,M_T=0)=\frac{1}{2}\left( Q_{{\Xi^0}{\Xi^-}}+Q_{{\Xi^-}{\Xi^0}}
\right) \\
\nonumber
\end{equation}
\begin{equation}
Q_{{\Xi}{\Xi}}(T=1,M_T=+1)=Q_{{\Xi^0}{\Xi^0}} \\
\nonumber
\end{equation}

{}From the above expressions it is easy to see that in isospin
saturated matter matter
(i.e. $k_{F}^{(n)}=k_{F}^{(p)}$, 
$k_{F}^{({\Sigma^+})}=k_{F}^{({\Sigma^0})}=k_{F}^{({\Sigma^-})}$ and
$k_{F}^{({\Xi^0})}=k_{F}^{({\Xi^-})}$)
the dependence on the third component of isospin disappears.

\newpage
\renewcommand{\theequation}{\Alph{section}.\arabic{equation}}
\setcounter{section}{2}
\section*{{\bf APPENDIX B: AVERAGE OF THE  CENTER--OF--MASS AND THE HOLE
MOMENTA}}

In this appendix we show how to compute an appropiate angular average of the
center--of--mass momentum of the pair $B_1B_2$ and the hole momentum
${\vec k}_{B_2}$ which enters in the determination of
the starting energy in Eqs.(\ref{eq:upot1}) and (\ref{eq:upot2}) . The center--of--mass
momentum $\vec{K}$ and
the
relative momentum $\vec{k}$ of the pair $B_1B_2$ are defined in the following
way:


\begin{equation}
  {\vec K} ={\vec k}_{B_1}+{\vec k}_{B_2} \ , 
\setcounter{equation}{1}
\end{equation}
\begin{equation}
 {\vec k} =\frac{M_{B_2}{\vec k}_{B_1}-M_{B_1}{\vec k}_{B_2}}{M_{B_1}+M_{B_2}}\,
\nonumber \\
= {\beta}{\vec k}_{B_1}-{\alpha}{\vec k}_{B_2} \ .
\end{equation}
From the above expressions it is easy to write ${\vec K}$ and ${\vec k}_{B_2}$ in
terms of the extrenal momentum $\vec{k}_{B_1}$ and the relative momentum $\vec{k}$, which is used as
integration variable in Eqs. (\ref{eq:upot1}) and (\ref{eq:upot2})
 

\begin{equation}
  {\vec K} =\frac{1}{\alpha}({\vec k}_{B_1}-{\vec k}) \ ,
\end{equation}
\begin{equation}
  {\vec k}_{B_2} =\frac{1}{\alpha}({\beta}{\vec k}_{B_1}-{\vec k}) \ . 
\end{equation}
The angle average of the center--of--mass momentum is defined as

\begin{eqnarray}
  \overline {K^{2}}(k_{B_1},k) &=& \frac{\displaystyle\int d(\cos{\theta})
K^{2}(k_{B_1},k,\cos{\theta})}{\displaystyle\int d(\cos{\theta})},
\end{eqnarray}
where
$K^{2}(k_{B_1},k,\cos{\theta})=\frac{1}{\alpha^{2}}(k_{B_1}^{2}
+k^{2}-2k_{B_1}k\cos{\theta})$, with $\theta$ being the angle between ${\vec k}_{B_1}$
and ${\vec k}$. The integration runs over all the angles for which
$|\vec{k}_{B_2}|<k_{F}^{(B_2)}$. Similarly, for the hole momentum we have
 
\begin{eqnarray}
  \overline {k^{2}_{B_2}}(k_{B_1},k) &=& \frac{\displaystyle\int d(\cos{\theta})
k^{2}_{B_2}(k_{B_1},k,\cos{\theta})}{\displaystyle\int d(\cos{\theta})},
\end{eqnarray}
where
$k^{2}_{B_2}(k_{B_1},k,\cos{\theta})=\frac{1}{\alpha^{2}}(\beta^{2}k_{B_1}^{2}
+k^{2}-2{\beta}k_{B_1}k\cos{\theta})$.

We can distinguish two cases in performing the angular integrals, $\beta$$k_{B_1}$ $<$
$\alpha$$k_{F}^{(B_2)}$   
and $\beta$$k_{B_1}$ $>$ $\alpha$$k_{F}^{(B_2)}$. In the first case, we have two possibilities: $0$
$<$ $k$ $<$ $\alpha$$k_{F}^{(B_2)}$$-$$\beta$$k_{B_1}$, for which all angle values are allowed,
giving the result
\begin{equation}
  \overline {K^{2}}(k_{B_1},k) = \frac{1}{\alpha^{2}}{\bigg [}k_{B_1}^{2}+k^{2}{\bigg ]}
\,
\end{equation}
\begin{equation}
  \overline {k_{B_2}^{2}}(k_{B_1},k) =
\frac{1}{\alpha^{2}}{\bigg [}\beta^{2}k_{B_1}^{2}+k^{2}{\bigg ]} \ ,
\end{equation}
and $\alpha$$k_{F}^{(B_2)}$$-$$\beta$$k_{B_1}$ $<$ $k$ $<$
$\alpha$$k_{F}^{(B_2)}$$+$$\beta$$k_{B_1}$, which have the following upper limit in the value of
$\cos{\theta}$
\begin{equation}
({\cos{\theta}})_{\rm max}=\frac{k^2+(\beta k_{B_1})^2-(\alpha k_{F}^{(B_2)})^2}{2{\beta}kk_{B_1}} \
,
\end{equation}
giving the result
\begin{equation}
  \overline {K^{2}}(k_{B_1},k) =
\frac{1}{\alpha^{2}}{\bigg 
[}k_{B_1}^{2}+k^{2}-\frac{1}{2\beta}{\bigg
(}({\beta}k_{B_1}+k)^{2}-({\alpha}k_{F}^{(B_2)})^{2}{\bigg )} {\bigg ]} \, 
\end{equation}
\begin{equation}
  \overline {k_{B_2}^{2}}(k_{B_1},k) =
\frac{1}{\alpha^{2}}{\bigg 
[}{\beta^{2}}k_{B_1}^{2}+k^{2}-\frac{1}{2}{\bigg 
(}({\beta}k_{B_1}+k)^{2}-({\alpha}k_{F}^{(B_2)})^{2}{\bigg )}{\bigg ]} \ . 
\end{equation}

In the second case, there is only one possibility: $\beta$$k_{B_1}$$-$$\alpha$$k_{F}^{(B_2)}$
$<$ $k$ $<$
$\alpha$$k_{F}^{(B_2)}$$+$$\beta$$k_{B_1}$ and the result is the same as in
the
previous case for the zone $\alpha$$k_{F}^{(B_2)}$$-$$\beta$$k_{B_1}$ $<$ $k$ $<$
$\alpha$$k_{F}^{(B_2)}$$+$$\beta$$k_{B_1}$. The result for the values $0$ $<$ $k$ $<$
$\beta$$k_{B_1}$$-$$\alpha$$k_{F}^{(B_2)}$ 
is zero because ${\vec k}_{B_2}$ is always larger than its Fermi sea.

This kind of average defines an angle--independent center--of--mass momentum and a hole momentum
(and therefore a starting energy)
for each pair $k_{B_1}$, $k$ so the angular integration in Eqs. (\ref{eq:upot1}) and
(\ref{eq:upot2}) can be performed analytically. Nevertheless, we still require to solve the
$G$--matrix equation for each pair of values $k_{B_1}$ and $k$, making the calculation much time
consuming. In order to speed up the procedure we introduce another average, which gives equivalent
results and saves a lot of time. For each external momentum $k_{B_1}$, we will only need to solve the
$G$--matrix equation for two values of the center--of--mas and hole momenta, which are obtained from
\begin{eqnarray}
  \overline {K^{2}}(k_{B_1}) &=& \frac{\displaystyle\int d^{3}k\,
K^{2}(k_{B_1},k,\cos{\theta})}{\displaystyle\int d^{3}k},
\label{aver1}
\end{eqnarray}
\begin{eqnarray}
  \overline {k^{2}_{B_2}}(k_{B_1}) &=& \frac{\displaystyle\int d^{3}k\,
k^{2}_{B_2}(k_{B_1},k,\cos{\theta})}{\displaystyle\int d^{3}k},
\label{aver2}
\end{eqnarray}
by limiting the integral over the modulus of $\vec{k}$ to the two possibilities mentioned above.
As before, we have the same cases $\beta$$k_{B_1}$ $<$ $\alpha$$k_{F}^{(B_2)}$ and $\beta$$k_{B_1}$
$>$
$\alpha$$k_{F}^{(B_2)}$ . Let's consider the first case. Now, when the integral over $k$ in Eqs.
(\ref{aver1}) and (\ref{aver2}) is limited to  $0$ $<$ $k$ $<$
$\alpha$$k_{F}^{(B_2)}$$-$$\beta$$k_{B_1}$ we have

\begin{equation}
  \overline {K^{2}}(k_{B_1}) =
\frac{1}{\alpha^{2}}{\bigg
[}k_{B_1}^{2}+\frac{3}{5}({\alpha}k_{F}^{(B_2)}-{\beta}k_{B_1})^{2}{\bigg ]} \ ,
\end{equation}
\begin{equation}
  \overline {k_{B_2}^{2}}(k_{B_1}) =
\frac{1}{\alpha^{2}}{\bigg 
[}{\beta}^{2}k_{B_1}^{2}+\frac{3}{5}({\alpha}k_{F}^{(B_2)}-{\beta}k_{B_1})^{2}{\bigg ]} \ ,
\end{equation}
whereas in the zone $\alpha$$k_{F}^{(B_2)}$$-$$\beta$$k_{B_1}$ $<$ $k$ $<$
$\alpha$$k_{F}^{(B_2)}$$+$$\beta$$k_{B_1}$ the expressions are a little bit more tedious

\begin{eqnarray}
 \overline {K^{2}}(k_{B_1}) &=&
{\bigg [}-\frac{\beta^{2}(1+{\beta}^{2})}{\alpha}k_{F}^{(B_2)}k_{B_1}^{4} 
+{\beta}(1+2{\beta}^{2}){k_{F}^{(B_2)}}^{2}k_{B_1}^{3} \, \nonumber \\
&+&{\bigg 
(}\frac{\alpha^{3}}{2}+\frac{\alpha}{12}(4{\beta}-26{\beta}^{2}-6){\bigg 
)}{k_{F}^{(B_2)}}^{3}k_{B_1}^{2} +{\alpha}^{2}{\beta}{k_{F}^{(B_2)}}^{4}k_{B_1} \, \nonumber \\ 
&+&\frac{\beta^{3}}{15{\alpha}^{2}}(5+3{\beta}^{2})k_{B_1}^{5}{\bigg ]} \, \nonumber \\
&\times&{\bigg [}({\alpha}k_{F}^{(B_2)})^{2}{\beta}k_{B_1}+\frac{1}{3}({\beta}k_{B_1})^{3} 
-{\alpha}k_{F}^{(B_2)}({\beta}k_{B_1})^{2}{\bigg ]}^{-1} \ ,  
\end{eqnarray}

\begin{eqnarray}
 \overline {k_{B_2}^{2}}(k_{B_1}) &=&
{\bigg [}-\frac{2{\beta}^{4}}{\alpha}k_{F}^{(B_2)}k_{B_1}^{4}
+3{\beta}^{3}{k_{F}^{(B_2)}}^{2}k_{B_1}^{3} \, \nonumber \\
&-&\frac{7}{3}{\alpha}{\beta}^{2}{k_{F}^{(B_2)}}^{3}k_{B_1}^{2} 
+{\alpha}^{2}{\beta}{k_{F}^{(B_2)}}^{4}k_{B_1} 
+\frac{8{\beta}^{5}}{15{\alpha}^{2}}k_{B_1}^{5}{\bigg ]} \, \nonumber \\
&\times&{\bigg [}({\alpha}k_{F}^{(B_2)})^{2}{\beta}k_{B_1}+\frac{1}{3}({\beta}k_{B_1})^{3} 
-{\alpha}k_{F}^{(B_2)}({\beta}k_{B_1})^{2}{\bigg ]}^{-1} \ . 
\end{eqnarray}
When ${\vec k}_{B_1}=0$ there exists only one zone of integration, $0$ $<$ $k$ $<$
$\alpha$$k_{F}^{(B_2)}$, and the average is very simple

\begin{eqnarray}
 \overline {K^{2}}(k_{B_1}) &=&
 \overline {k_{B_2}^{2}}(k_{B_1})=\frac{3}{5}{k_{F}^{(B_2)}}^{2} \ . 
\end{eqnarray} 

Finally, in the second case, $\beta$$k_{B_1}$ $>$ $\alpha$$k_{F}^{(B_2)}$, there is also
only one
integration zone, 
$\beta$$k_{B_1}$$-$$\alpha$$k_{F}^{(B_2)}$$<$ $k$ $<$$\alpha$$k_{F}^{(B_2)}$$+$$\beta$$k_{B_1}$, 
and the corresponding averages are

\begin{eqnarray}
 \overline {K^{2}}(k_{B_1}) &=&
\frac{3}{5}{k_{F}^{(B_2)}}^{2}+k_{B_1}^{2} \,  
\end{eqnarray}

\begin{eqnarray}
 \overline {k_{B_2}^{2}}(k_{B_1}) &=&
\frac{3}{5}{k_{F}^{(B_2)}}^{2} \ . 
\end{eqnarray}


\vfil
\newpage
 
\begin{figure}
       \setlength{\unitlength}{1mm}
       \begin{picture}(100,180)
       \put(15,10){\epsfxsize=12cm \epsfbox{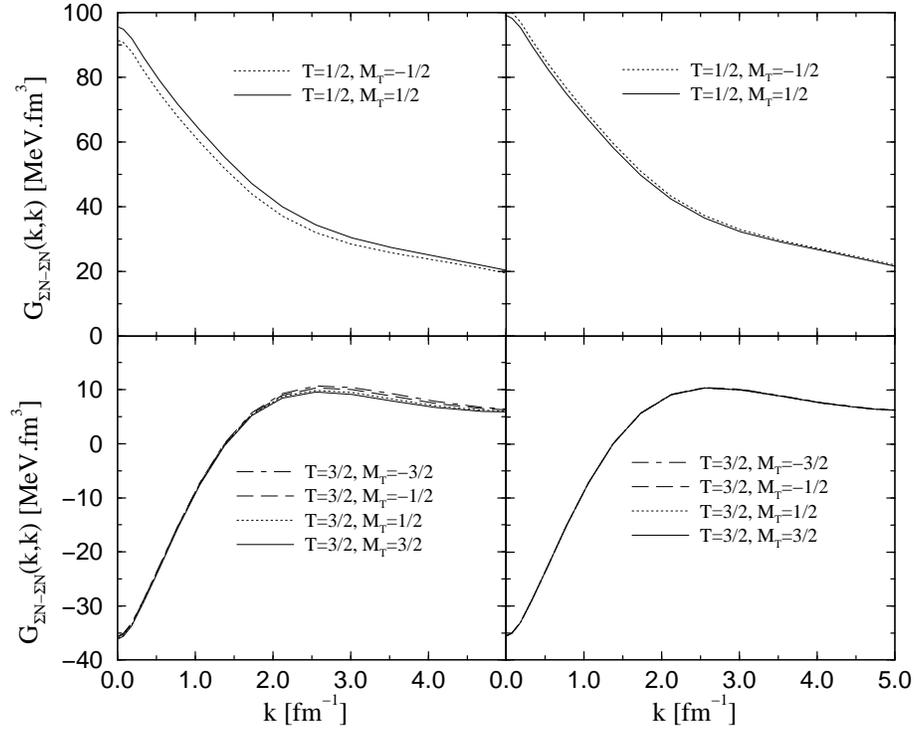}}
       \end{picture}
\caption{Diagonal $\Sigma N$ G--matrix in the $^{1}S_{0}$ partial wave as a function of the relative
momentum at a density $\rho=0.6$ fm$^{-3}$, for the different $(T,M_T)$ isospin channels. The right
panels are for symmetric nuclear matter, $x_n=x_p=0.5x_N$, while the left ones correspond to
$x_n=3 x_p=0.75x_N$. In all cases $x_{\Sigma^-}=0.1$ and $x_{\Lambda}=0$.} 
\label{fig:gmatrix}
\end{figure}

\begin{thebibliography}{200}


\bibitem{glendenning92}   N.~K.\ Glendenning, Phys.\ Rev.\ D {\bf
46}, 1274 (1992).
\bibitem{lattimer91} J. Lattimer, C. Pethick, M. Prakash and P.
Haensel,
Phys. Rev. Lett. {\bf 66}, 2701 (1991).
\bibitem{cook94} G.B. Cook, S.L. Shapiro and S.A. Teukolsky,
Astrophys. J {\bf
424}, 823 (1994). 
\bibitem{pke95} R.\ Knorren, M.\ Prakash and P.~J.\ Ellis,
Phys.\ Rev.\ C {\bf 52}, 3470 (1995).
\bibitem{prakash97} M.\ Prakash, I.\ Bombaci, M.\ Prakash,
                    P.~J.\ Ellis, J.M.\ Lattimer, and R.\
Knorren,
                    Phys.\ Rep.\ {\bf 280}, 1 (1997).
\bibitem{ellis95} P.J. Ellis, R. Knorren and M. Prakash, Phys.
Lett. {\bf B349},
11 (1995).
\bibitem{ms96} J.\ Schaffner and I.\ Mishustin, Phys.\ Rev.\
C {\bf 53}, 1416 (1996).
\bibitem{bg97} S.~Balberg and A.~ Gal, Nucl.\ Phys.\ {\bf A625},
435 (1997).
\bibitem{schulze1} H.-J.\ Schulze, A.\ Lejeune, J. Cugnon, M.\
Baldo, and 
U.\ Lombardo, Phys. Lett. {\bf B355}, 21 (1995).
\bibitem{schulze2} H.-J. Schulze, M. Baldo, U. Lombardo, J.
Cugnon, 
A. Lejeune, Phys. Rev. {\bf C57}, 704 (1998).
\bibitem{schulze3}
M.\ Baldo, G.F.\ Burgio, and H.-J.\ Schulze, Phys.\ Rev.\ C {\bf
58}, 3688 (1998).
\bibitem{sr99} V.~G.~J.\ Stoks and Th.~A.\ Rijken,
Phys.\ Rev.\ C {\bf 59}, 3009 (1999)
\bibitem{sl99} V.~G.~J.\ Stoks and T.-S.~H.\ Lee, Phys.\ Rev.\ C {\bf
60}, 024006 (1999).
\bibitem{bombaci91}
I. Bombaci and U. Lombardo, Phys. Rev. C {\bf 44}, 1892 (1991)
\bibitem{engvik94}
L. Engvik, M. Hjorth-Jensen, E. Osnes, G. Bao and E. {\O}stgaard,
Phys. Rev. Lett. {\bf 73},
2650 (1994)
\bibitem{engvik96}
L. Engvik, M. Hjorth-Jensen, E. Osnes, G. Bao and E. {\O}stgaard,
Astrophys. J. {\bf 469}, 794 (1996).

\bibitem{rsy98} Th.~A.\ Rijken, V.~G.~J.\ Stoks, and Y.\
Yamamoto, Phys.\ Rev.\ C {\bf 59},
                21 (1998).
\bibitem{isaac99}
I.\ Vida\~na, A.\ Polls, A.\ Ramos, L.\ Engvik, and M.\ 
Hjorth-Jensen, in preparation.
\bibitem{nijme89} P.M.M. Maesen, Th.~A.\ Rijken, and J.J. Swart, 
 Phys.\ Rev.\ C {\bf 40},
                2226 (1989).
\bibitem{bmz90} H.~Band\={o}, T. ~Motoba, and J. Zofka, Int. J.
Mod. Phys. 
{\bf A5}, 4021 (1990).
\bibitem{morten96} M. Hjorth-Jensen, A. Polls, A. Ramos, and H.
M\"uther, 
Nucl. Phys. {\bf A605},458 (1996).
\bibitem{isa98} I. Vida\~na, A. Polls, A. Ramos, and M.
Hjorth-Jensen,
Nucl. Phys. {\bf A644}, 201 (1998).
\bibitem{yama94}
Y. Yamamoto, T. Motoba, H. Himeno, K. Ikeda and S. Nagata,
Prog. Theor. Phys. Suppl. {\bf 117}, 36 (1994).
\bibitem{MDG}
D.J. Millener, C.B. Dover and A. Gal, Phys. Rev. C {\bf 38}, 2700 (1988).




\end{thebibliography}
\end{document}